\documentclass[aps,twocolumn,showpacs,amsmath,amssymb,prl
]{revtex4-1}

\usepackage[pdftex]{graphicx}
\usepackage[usenames,dvipsnames]{color}
\usepackage{epstopdf}
\usepackage[hidelinks]{hyperref}

\newcommand{\Nloop}[1]{\mathcal{N}}
\newcommand{\Jloops}[1]{\mathcal{J}}

\newcommand{\N}[1]{\Omega}
\newcommand{\Len}[1]{L}

\begin{document}
\title{Selecting Initial States from Genetic Tempering for Efficient Monte Carlo Sampling}
\author{Thomas E. Baker}
\affiliation{Institut quantique \& D\'epartement de physique, Universit\'e de Sherbrooke, Qu\'ebec, Canada J1K 2R1}
\date{\today}

\begin{abstract}
An alternative to Monte Carlo techniques requiring large sampling times is presented here.  Ideas from a genetic algorithm are used to select the best initial states from many independent, parallel Metropolis-Hastings iterations that are run on a single graphics processing unit.   This algorithm represents the idealized limit of the parallel tempering method and, if the threads are selected perfectly, this algorithm converges without any Monte Carlo iterations--although some are required in practice.  Models tested here (Ising, anti-ferromagnetic Kagome, and random-bond Ising) are sampled on a time scale of seconds and with a small uncertainty that is free from auto-correlation.
\end{abstract}

\maketitle

\paragraph{Introduction} 
Monte Carlo (MC) sampling is a ubiquitous technique with many applications, such as sampling spin systems and integration \cite{landau1979monte}. The canonical implementation is the Metropolis-Hasting (MH) algorithm \cite{metropolis1953equation,hastings1970monte}, but there are a host of additions that one can add for improvement \cite{landau1979monte,wang2001efficient}.  A large number of MC samples are required to generate good statistics, so care must be taken in implementing any MC algorithm to ensure that it scales to larger systems.  Even with an efficient implementation, the algorithm can still take a long time to run. Also, the possibility exists that results become biased due to small errors in random number generation and also that successive samples are auto-correlated \cite{landau1979monte}.

\begin{figure}[b]
\begin{centering}
\includegraphics[width=0.95\columnwidth]{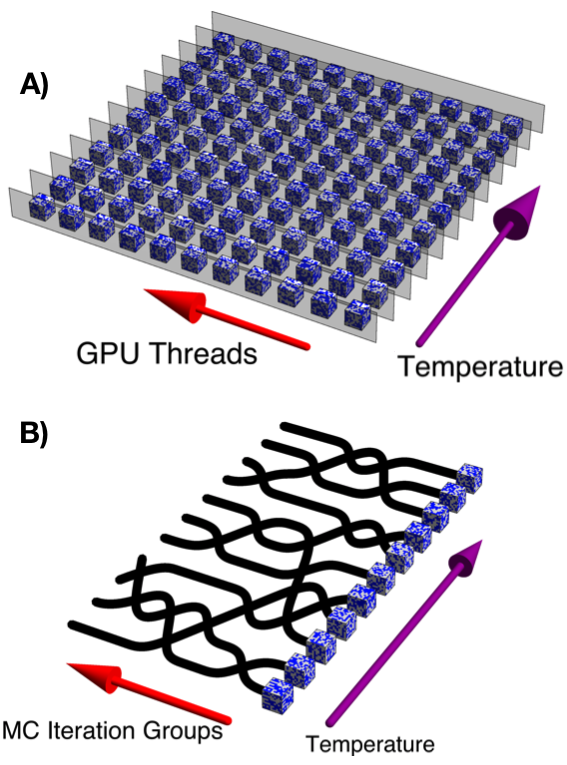}
\end{centering}
\caption{(color online) A visualization of the comparison between (A) GT-GPU and (B) PT \cite{swendsen1986replica}.  Each cube represents a spin system of arbitrary dimension.  The threads on a GPU (one per cube in A) are passed to the next temperature (across gray separators) and effectively sample every possible crossing in PT (black lines in B). Each cube in GT-GPU is independent from the others at the same temperature, corresponding to perfectly uncorrelated MC iteration groups in PT.\label{GPUvPT}}
\end{figure}

It is sometimes the case that improved hardware can outperform even the most clever algorithms.  In the case of MC sampling, a natural place to look for that improvement is by increasing the number of computer cores for parallel computation.  The only limitation of this strategy is the physical computing resources available.  In reality, the cost and availability of central processing unit (CPU) cores to the user will set the limit of this parallelization.

Alternatively, a graphics processing unit (GPU) is available on nearly all computers and has potentially hundreds of thousands or more of cores for simple operations.   Previous uses of GPUs add extra details to the traditional MH algorithm (modified updates, partitioning schemes, cluster algorithms, intricate GPU implementation, etc. \cite{preis2009gpu,lee2010utility,preis2009gpu,navarro2016adaptive,beam2016fast,block2010multi,isakov2015optimised}) with the intention of making each MC step as fast as possible.  Another possible way to use a GPU can be explored.

This leads to the central question of this work: is using multiple threads on a GPU with only MH as good as using the advanced techniques on a CPU for MC? The answer is yes.  Traditional MH parallelized on a single, commonly available GPU (using each thread as an independent MH algorithm and selected as in a genetic algorithm) gives results that rival other methods quickly.

The proposed method, genetic algorithm tempering  on a GPU (GT-GPU), is summarized in Fig.~\ref{GPUvPT}.  The threads which are kept and propagated are selected to reduce systematic errors from insufficient relaxation of states to the target ensemble. Creating $\Jloops{}$ independent threads at a temperature (split by gray dividers in Fig.~\ref{GPUvPT}) on a GPU of simultaneous, uncorrelated MH iterations will sample every crossing available in a parallel tempering (PT) algorithm \cite{swendsen1986replica}.  PT resembles a renormalization group algorithm, passing information between high and low temperatures.  GT-GPU represents the idealized limit of PT and achieves ergodicity faster \cite{neirotti2000approach}.   Also, many minima are sampled simultaneously in GT-GPU, circumventing large energy barriers that might separate minima.

The order of the paper is as follows: First, GT-GPU is explained and shown to be highly accurate due to selection of threads from inside of a hysteresis loop--obtained from partially unrelaxed samples. Results for several lattice models (Ising, Kagome \cite{takagi1993magnetic}, and random-bond Ising Model (RBIM) \cite{honecker2001universality}) are shown.  An analysis of the uncertainty shows why GT-GPU achieves high precision in comparison with single-thread algorithms.  Then, phase transitions are shown.  A 400 line code is provided \cite{code}.

\paragraph{Spin Hamiltonians with Metropolis-Hastings} 
The general class of Hamiltonian considered here is
\begin{equation}\label{isingHam}
\mathcal{H} = -\sum_{\langle ij\rangle}JS^z_iS^z_j
\end{equation}
where the spin operators $S^z$ represent classical half spins on each of $\N{}=L^D$ sites (dimension $D$) indexed by $i$ and $j$.  The interaction is taken over the nearest neighbors, denoted by $\langle\rangle$.  The algorithm described in the following applies to any open or closed boundary condition, any $D$, coupling $J$ (or $J_{ij}$), $n^\mathrm{th}$ order interactions, non-square and non-cubic lattices, different spin magnitudes, etc.

To sample Eq.~\eqref{isingHam}, MH iterations are taken: \cite{metropolis1953equation,hastings1970monte}
\begin{enumerate}
\item Choose a lattice site $i$ at random.
\item Find the energy difference $\delta E$ if the spin is flipped.
\item If $\delta E<0$, flip the spin.
\item If the previous point fails, evaluate for some temperature $T$ and random number $q$ the expression
\begin{equation}\label{flipcondition}
-T\log(q)>\delta E
\end{equation}
and flip the spin if the condition is true.
\end{enumerate}

\paragraph{Graphics processing units}
The MH algorithm only requires IF and FOR statements to be executed.  A GPU thread can execute both, so each thread can execute an independent set of single-loop MH iterations.   In all, this algorithm requires the generation of two random numbers per iteration (for $i$ and $q$).  It is possible to generate random numbers on the GPU, but this can be memory intensive.  Note that the number of threads on a GPU is not a fixed hardware feature.  The number of independent MH iterations is limited by memory.  Sacrificing `on-the-fly' random number generation means new random numbers are buffered into the GPU over $\eta$ cycles.

GPUs run much faster when using only single-precision (32-bit) or less.  Double-precision, the standard for scientific application, can slow the GPU significantly.  For many GPUs, double-precision is not even possible. Only single precision is used on the GPU.  To minimize the impact of using the single-precision (and prevent overflows), the sum of quantities of interest is returned to the CPU after each cycle to be proccessed with double precision.  Only Eq.~\eqref{flipcondition} is affected by single-precision.
\begin{figure}[b]
\includegraphics[width=\columnwidth]{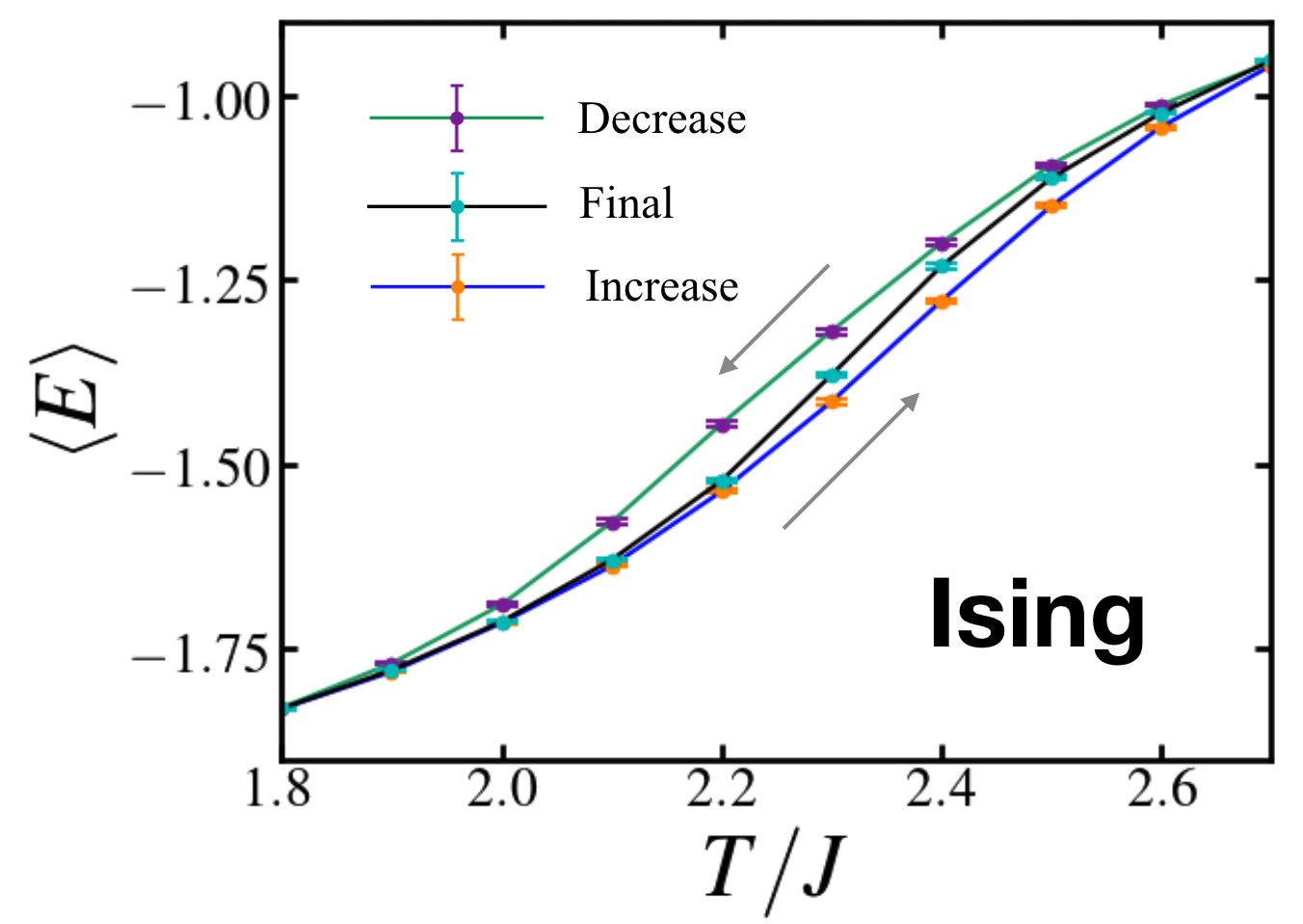}
\caption{\label{hyst} A hysteresis loop where final states are passed from higher (lower) temperatures as the temperature decreases (increases) is plotted for the periodic $D=2$ Ising model evaluated with $L=16$ and $\Jloops{}=1000$, and $\eta\Nloop{}=10^3$ for the hysteresis loop and 10 times more for the final run.  A delay of $\eta\Nloop{}=5\times10^3$ was conducted before taking all samples.  The entire run between $T/J=0.5$ and 5.0 with $\Delta T=0.1$ was 80 seconds on a single 2.8 GHz processor with one GPU.
}
\end{figure}

\paragraph{Genetic Tempering}
The algorithm is presented in two steps.  First, standard MC over $\Jloops{}$ threads at each $T$--initialized from the previous $T$--is run.  The average on thread $w$ of quantity $x$ ($E$, $M$, $|M|$, $M^2$, $M^4$, etc. where $M_t=\N{}^{-1}\sum_{i=1}^{\N{}}(S_i^z)_{(t)}$) is
\begin{equation}\label{singleAvg}
\overline x^{(w)} = \frac1{\Nloop{}^{(w)}}\sum_{t=1}^{\Nloop{}^{(w)}}x_t^{(w)}
\end{equation}
where $\Nloop{}^{(w)}$ is the number of samples encountered on a thread $w$. The average over all threads is
\begin{equation}\label{threadAvg}
\langle x\rangle = \frac1{\Jloops{}}\sum_{w=1}^{\Jloops{}}\overline{x}^{(w)}
\end{equation}
and the uncertainty is the standard deviation of the mean
\begin{equation}\label{stdReg}
\sigma_\mathrm{GPU}=\frac{\sigma(\Jloops{})}{\sqrt{\Jloops{}}}=\sqrt{\frac{\sum_{w=1}^{\Jloops{}}\left(\overline{x}^{(w)}-\langle x\rangle\right)^2}{\Jloops{}(\Jloops{}-1)}}
\end{equation}
where $\Nloop{}^{(w)}\rightarrow\Nloop{}_\mathrm{GPU}^{(w)}$ in Eq.~\eqref{threadAvg}. The final wavefunction and $\langle E\rangle$ for the sweep are saved on the CPU.  The results from increasing, $T_>$, and decreasing, $T_<$, the temperature form a hysteresis loop, shown in Fig.~\ref{hyst}, since the samples are not fully converged.  These sweeps are not run for long, $t\ll\tau$ for some relaxation time $\tau$; however, they are precise (justified formally after this section).

An understanding of the relaxation time is necessary obtain accuracy.  The decay function, $K(t)$, relates initial ($t=0$) and final ensembles ($t\rightarrow\infty$) \cite{landau1979monte}
\begin{equation}\label{relaxation}
\langle x(0)-x(\infty)\rangle K(t)=\langle x(t)-x(\infty)\rangle
\end{equation}
where $\langle x(z)\rangle$ is the same as Eq.~\eqref{singleAvg} up to time $t=z$. The factor $K(t)$ is known to be $\exp(-(t/\tau)^\nu)$ in the long time limit for arbitrary models with $\nu\leq1$ and $\nu=1$ capturing many systems \cite{landau1979monte}.  The exponential nature can be loosely justified on the Boltzman weight from Eq.~\eqref{flipcondition} ({\it i.e.}, an excitation has an exponential probability to exist and therefore a related number of time steps to reach).  The average decay of an ensemble is best represented as
\begin{equation}\label{avgDecay}
K^\mathrm{GT}(t)=\exp(-(t/\tau)^\nu+\ln|\langle x(0)-x(\infty)\rangle/\langle x(\infty)\rangle |)
\end{equation}
after dividing Eq.~\eqref{relaxation} by $\langle x(\infty)\rangle$.  The logarithmic term in the argument of Eq.~\eqref{avgDecay} shows that if the proper threads are selected ($\langle x(0)\rangle\approx\langle x(\infty)\rangle$), the algorithm would converge automatically at $t=0$, independent of $\nu$ and $\tau$.   So, to achieve accuracy, only selected threads should be kept to ensure the initial ensemble is centered on the final.  

The second step of the algorithm therefore selects initial states between $\langle x\rangle_{T_>}$ and $\langle x\rangle_{T_<}$ (presented here is $x=E$) at random from the saved states.  A series of delay cycles can be run before taking data.  The final, third sweep is run and has vastly improved accuracy.  If $t=0$, this implies the hysteresis states are averaged and this may not always be the accurate answer in the realistic case of imperfect selection.  Thus, samples are propagated for a time after selection which re-weights the states towards the target answer.  The samples do not need to be fully converged for an accurate average.

\begin{figure}[b]
\includegraphics[width=\columnwidth]{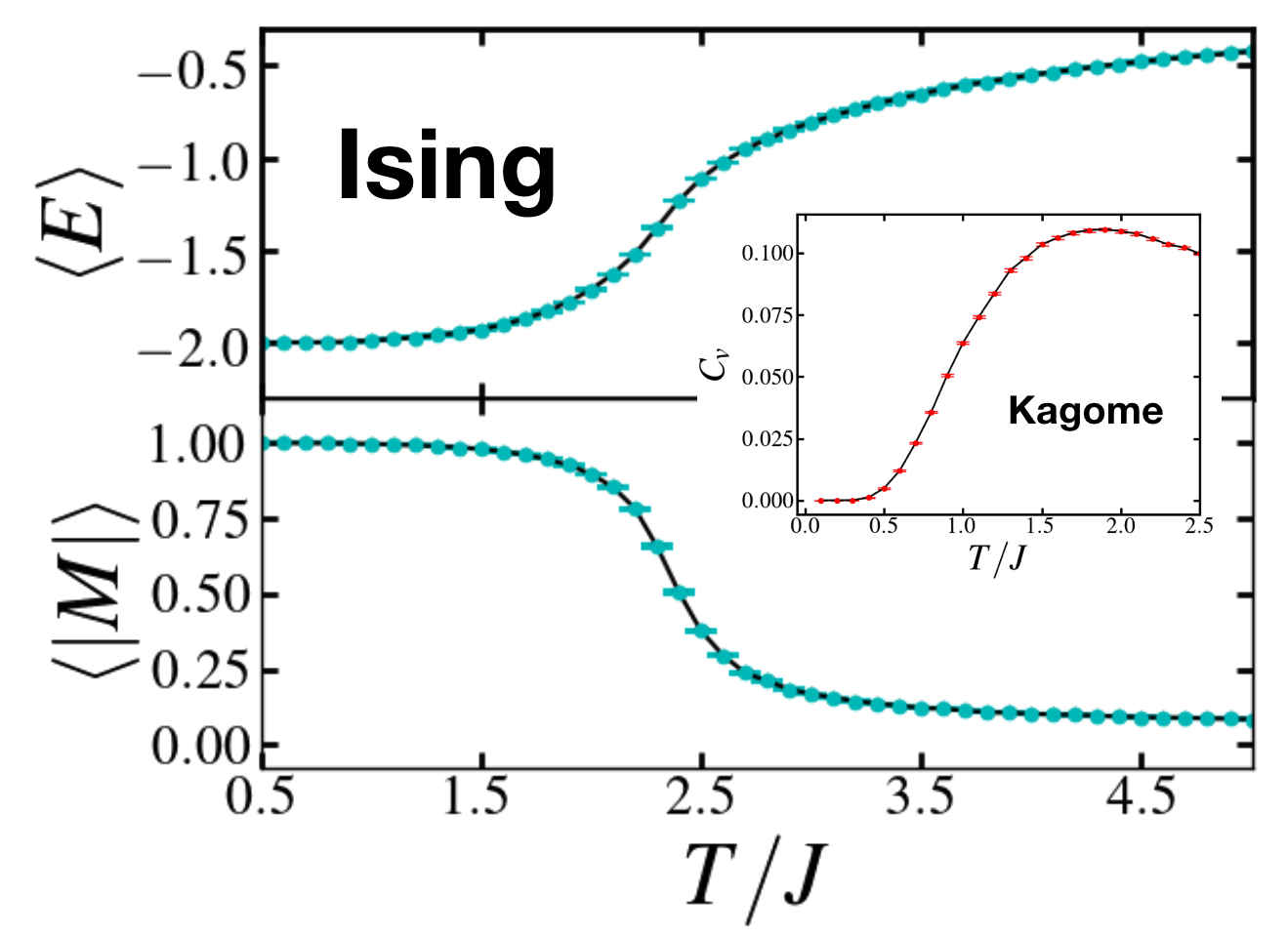}
\caption{The full, final data from Fig.~\ref{hyst} for the Ising model.  The inset shows the specific heat per site for the Kagome lattice with anti-ferromagnetic coupling, matching Ref.~\onlinecite{takagi1993magnetic}, which took 90 seconds under the same parameters as Fig.~\ref{hyst}.
\label{Ising_EM}}
\end{figure}

\paragraph{Ising model results}
Results for the $D=2$ periodic, ferromagnetic Ising model are shown in Fig.~\ref{Ising_EM}. Each uncertainty is approximately $10^{-3}$ or less and the data was generated in 80 seconds.  The inset of Fig.~\ref{Ising_EM} shows the specific heat per site ($C_v$)
\begin{equation}\label{specificheat}
C_v=\N{}(\langle E^2\rangle-\langle E\rangle^2)/T^2
\end{equation}
of the anti-ferromagnetic Kagome lattice, matching Ref.~\onlinecite{takagi1993magnetic} and taking 90 seconds to generate.

\paragraph{Improved precision} 
Note that the hysteresis loop in Fig.~\ref{hyst} has very small error bars.  These signify precision.  Precision does not mean the value is free from systematic error.  For example, $\langle M\rangle$ could have been plotted in Fig.~\ref{Ising_EM} and noticed to be very smoothly decaying from value 1 at $T=0$ (traditional MH would produce large oscillations \cite{landau1979monte}) if an ordered state is initialized.  Yet, the true value is zero everywhere.  This illustrates that the error bar on the hysteresis loop of Fig.~\ref{Ising_EM} is the precision between runs. The gap between $\langle x\rangle_{T_<}$ and $\langle x\rangle_{T_>}$ is the systematic error from insufficient relaxation, corrected in the last sweep. 

The source of the precision over single thread MC is discussed here. Consider that the limit of large statistics (denoted $\mathrm{Avg}$) of the summation in Eq.~\eqref{stdReg} as
\begin{equation}\label{bigAvg}
\mathrm{Avg}\left(\overline{x}^{(w)}-\langle x\rangle\right)=0\pm\Xi{\Big/}{\sqrt{\Nloop{}_\mathrm{GPU}}}
\end{equation} 
by definition of the standard deviation of the mean, $\Nloop{}_\mathrm{GPU}$ is averaged over all $w$, and $\Xi$ is the true uncertainty of $\overline x$.  The usefulness of Eq.~\eqref{bigAvg} is it represents the expected result of many runs of the GT-GPU algorithm.  From the mean value theorem, the summation in Eq.~\eqref{stdReg} is effectively replaced by a sum of Eq.~\eqref{bigAvg}, giving
\begin{equation}\label{GPUAVG}
\mathrm{Avg}\left(\sigma_\mathrm{GPU}\right)=\Xi{\Big/}\sqrt{\Nloop{}_\mathrm{GPU}(\Jloops{}-1)}
\end{equation}
for the expected GT-GPU uncertainty of many runs.

Contrastingly, the uncertainty in a single MC loop, $\sigma_\mathrm{MC}$, is known to be $\sigma_\mathrm{MC}=\sigma(\Nloop{}_\mathrm{MC})\sqrt{(1+2\tau_\mathrm{AC})/\Nloop{}_\mathrm{MC}}$ for some auto-correlation time $\tau_\mathrm{AC}$ \cite{landau1979monte}.  The realistic limit where this time will be run is for $\Nloop{}_\mathrm{MC}\gg\tau_\mathrm{AC}$ away from a critical temperature, $T_c$.  The appropriate expression for single-thread MC is therefore $\mathrm{Avg}\left(\sigma_\mathrm{MC}\right)\gtrsim\Xi/\sqrt{\Nloop{}_\mathrm{MC}}$. The resulting ratio of averaged standard deviations is 
\begin{equation}\label{compareError}
\frac{\mathrm{Avg}\left(\sigma_\mathrm{GPU}\right)}{\mathrm{Avg}\left(\sigma_\mathrm{MC}\right)}
\lesssim\sqrt{\frac{\Nloop{}_\mathrm{MC}}{\Jloops{}\Nloop{}_\mathrm{GPU}}}
\end{equation}
where $(\Nloop{}_\mathrm{MC},\Nloop{}_\mathrm{GPU},\Jloops{})\gg1$ is the limit where the statistical formulas, Eq.~\eqref{stdReg}, apply.  The inequality expresses that $\tau_\mathrm{AC}$ is not taken into account and that the estimation of the statistics is approximate.  Note that GT-GPU will be much more precise near $T_c$ in any case since $\tau_\mathrm{AC}$ is not in Eq.~\eqref{stdReg}.  Take further note that this expression shows that $\Jloops{}$ and $\Nloop{}_\mathrm{GPU}$ contribute equally to decreasing the uncertainty. For example, $\Jloops{}=100$ and $\Nloop{}_\mathrm{GPU}=10^4$ gives an order of magnitude uncertainty similar to $\Nloop{}_\mathrm{MC}=10^6$ away from $T_c$, and better near it.

\begin{figure}
\includegraphics[width=\columnwidth]{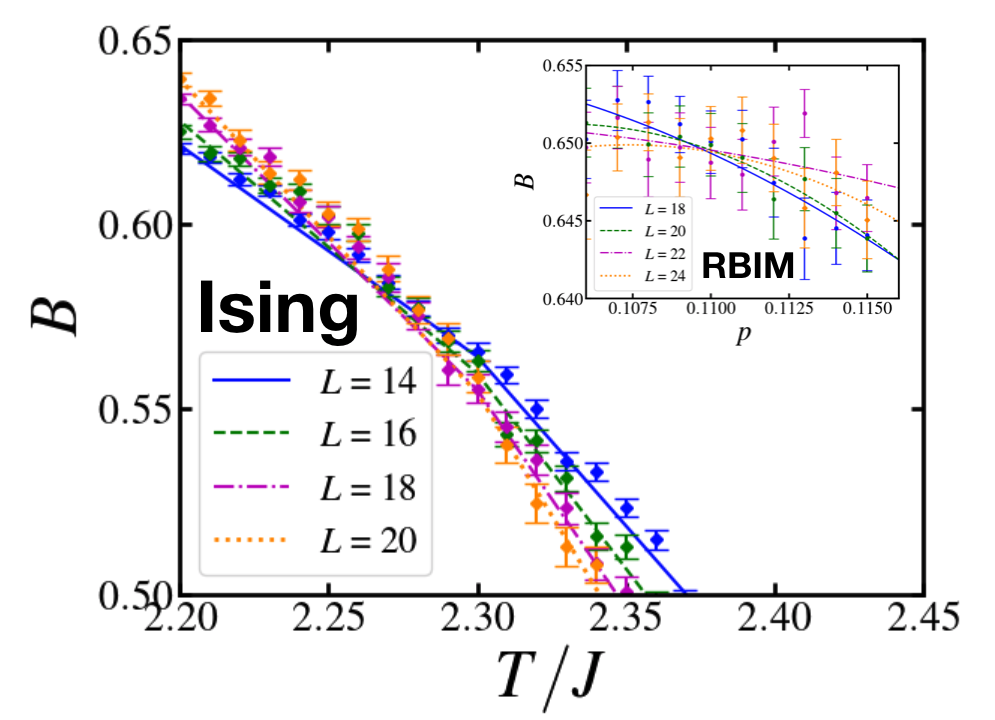}
\caption{\label{PhaseTransition} Data points and quadratic best fit lines for $B$ \{inset\} for $L=14,16,18, 20$ \{18, 20, 22, 24\} \{random-bond\} Ising model.  The parameters are $\Jloops{}=10^3$, $\eta\Nloop{}=10^5/10^3$ $\{5\times10^{5}/10^{3}\}$ and $10^4/(5\times10^4)$ $\{10^{5}/(5\times10^{5})\}$ for the final and initial sweeps (delay/runs) taking 20 \{15\} minutes per $L$.
} 
\end{figure}

\paragraph{Phase Transitions} 
The Ising model's Binder cumulant $B$ (shown in Fig.~\ref{PhaseTransition}) is
\begin{equation}\label{binder}
B=1-\langle M^4\rangle/(3\langle M^2\rangle^2)
\end{equation}
which is dimensionless and has a crossing at $T_c$ for various $L$.  The hysteresis loop does not need to be generated with a uniform temperature step everywhere.  A coarse calculation can reveal ({\it i.e.}, by identifying a peak in $C_v$) that the region of interest for the Ising model is $T/J\in[2.2,2.4]$ with a resolution of $\Delta T/J=0.1$.  Inside of the region of interest a sweep with $\delta T/J=10^{-2}$ is run.  A small $\delta T$ is useful since $\langle x(0)\rangle_T\approx\langle x(\infty)\rangle_{T+\delta T}$ as $\delta T\rightarrow0$.  Minimizing over the difference in the fits squared gives $T_c^\mathrm{GT}=2.267\pm0.005$ ($T_c^\mathrm{Ising}=2/\ln(1+\sqrt2)\approx2.269$ \cite{landau1979monte}).

\paragraph{Systems with Disorder}
The possibility to assign different configurations to each GPU thread can be explored by sampling the RBIM where $J\rightarrow J_{ij}$ with value $\pm1$ in Eq.~\eqref{isingHam} \cite{honecker2001universality}.  A set (one per $w$)  of random numbers $[0,1]$ for each bond are stored and compared with a number $p$ ($p\in[0.01,0.20]$ with $\Delta p=0.01$).  If the stored number is less than $p$, $J_{ij}=-1$ and $J_{ij}=1$ otherwise.  Nishimori's condition is used, $1-p=p\exp(2/T)$ \cite{honecker2001universality}.  The final sweep was initialized randomly between the increasing and decreasing sweeps.  Data was run over $p\in[0.01,0.20]$ with $\Delta p=10^{-2}$, focused on $[0.09,0.12]$ with $\delta p=10^{-3}$.  Note $p_c^\mathrm{GT}=0.1097\pm5\times10^{-4}$ (Ref.~\onlinecite{honecker2001universality}: $p_c\approx0.1094\pm2\times10^{-4}$).  The RBIM was chosen to run for $\Nloop{}$ to nearly match the Ising model's time to focus on thread selection, not $\Nloop{}$. 

\paragraph{Conclusions} 
The standard Metropolis-Hastings algorithm is implemented on separate, independent GPU threads. GT-GPU achieves great precision in a small amount of time due to sampling different minima separately.  High accuracy can be obtained by choosing selected threads to propagate, in this case from random sampling inside of a hysteresis loop generated from partially relaxed samples. Models with and without disorder and frustration were sampled very accurately. This is a great departure from other methods relying on many MC iterations; GT-GPU uses initial configurations of threads to obtain ensemble statistics instead.  Methods to select threads are encouraged to be developed going forward. 

\paragraph{Acknowledgements} 
Funding for this project was provided solely by the postdoctoral fellowship from Institut quantique. This research was undertaken thanks in part to funding from the Canada First Research Excellence Fund (CFREF). This research was enabled in part by support provided by Calcul Qu\'ebec (www.calculquebec.ca) and Compute Canada (www.computecanada.ca).  Computations were made on the supercomputer Helios from Universit\'e Laval (with funding also by Universit\'e de Montreal), managed by Calcul Qu\'ebec and Compute Canada. The operation of this supercomputer is funded by the Canada Foundation for Innovation (CFI), the ministère de l'Économie, de la science et de l'innovation du Qu\'ebec (MESI) and the Fonds de recherche du Qu\'ebec - Nature et technologies (FRQ-NT).  Computations were also made on the supercomputer Graham from the University of Waterloo, managed by Compute Canada.  The author also appreciates conversation with Jeanne Colbois, Colin Trout, Yehua Liu, David Aceituno, Benjamin Brown, Maxime Charlebois, Benjamin Bourassa, Prashanth Jaikumar, Glen Evenbly, Andr\'e-Marie Tremblay, and David Poulin.

\bibliography{MCgpu}

\end{document}